\documentstyle[emulateapj]{article}
\received{}
\revised{}
\accepted{}
\journalid{}{}
\articleid{}{}
\paperid{}
\cpright{}{}
\ccc{}
\lefthead{Matthews \& Gao}
\righthead{CO Detections of LSB Galaxies}
\begin{document}

\newcommand{\msun}{\mbox{${\cal M}_\odot$}}
\newcommand{\lsun}{\mbox{${\cal L}_\odot$}}
\newcommand{\kms}{\mbox{km s$^{-1}$}}
\newcommand{\HI}{\mbox{H\,{\sc i}}}
\newcommand{\mhi}{\mbox{${\cal M}_{HI}$}}
\def\hst{{\it HST}}
\def\etal{{\it et al.}}
\newcommand{\HII}{\mbox{H\,{\sc ii}}}
\newcommand{\am}[2]{$#1'\,\hspace{-1.7mm}.\hspace{.0mm}#2$}
\newcommand{\as}[2]{$#1''\,\hspace{-1.7mm}.\hspace{.0mm}#2$}
\newcommand{\dark}{$M_{HI}/L_{B}$}
\newcommand{\grat}{$M_{H2}/M_{HI}$}
\newcommand{\conv}{CO-to-H$_{2}$}

\submitted{Accepted to ApJ Letters}
\title{CO Detections of Edge-On Low Surface Brightness Galaxies}

\vskip0.5cm
\author{L. D. Matthews\altaffilmark{1}}
\author{Yu Gao\altaffilmark{2}}

\altaffiltext{1}{National Radio Astronomy Observatory, 520 Edgemont Road,
Charlottesville, VA 22903 USA,
Electronic mail: lmatthew@nrao.edu}
\altaffiltext{2}{Infrared Processing and Analysis Center, MS 100-22, Caltech,
Pasadena, CA 91125}

\singlespace
\tighten
\begin{abstract}
We have obtained deep CO(1-0) observations of 8
nearby, edge-on, low surface brightness (LSB) spiral galaxies using
the NRAO 12-m telescope. We report detections of 3 of our targets
at $>4\sigma$ level as well as one marginal ($\sim3\sigma$)
detection. These are among the first direct detections of molecular
gas in late-type, LSB spirals. Using
a Galactic CO-to-H$_{2}$ conversion factor, we derive
$M_{H2}\sim 1.4-3.5\times 10^{7}$~\msun\ and \grat$\sim$0.010-0.055 for
the detected sources, with 3$\sigma$ upper limits of
$M_{H2}\la0.7-5.7\times10^{7}$~\msun\ and \grat$\la$0.006-0.029 for the
undetected objects.

\end{abstract}


\section{Introduction}
Low surface brightness (LSB) disk
galaxies can be defined as rotationally-dominated 
galaxies with extrapolated central surface
brightnesses $\mu_{B}(0)\ga$23~mag arcsec$^{-2}$.   
Numerous studies have
now shown that these are not abnormal objects, but rather are
a common product of disk galaxy formation and evolution. 

Compared with disk
galaxies of high surface brightness (HSB),  most late-type
LSBs tend to be  \HI-rich, with diffuse, 
low-density stellar disks,  and
metallicities $\la$1/3 solar (e.g., McGaugh 1994). 
This suggests that  LSB galaxies have
been inefficient star-formers over their lifetimes. 
Indeed, 
\HI\ surface densities in LSB galaxies are often observed to be
well below the critical threshold for
star formation (e.g., van der Hulst et al. 1993). Nonetheless,
LSB galaxies typically contain some
signatures of ongoing star formation, including modest amounts of
H$\alpha$ emission (e.g., Schombert et al. 1992) 
and blue colors indicative of young stellar populations (e.g.,  McGaugh
\& Bothun 1994; Matthews \& Gallagher 1997). 

In spite of the evidence for modest ongoing star formation, 
a number of observational and 
theoretical arguments suggest
that LSB galaxies may be extremely molecular gas-deficient 
systems. For example,
LSB galaxies are generally found to be dust-poor 
(e.g., Matthews \& Wood 2001; hereafter MW01). This, combined with possible
low interstellar pressures, and the lack of the cooling effects of
metals, might be expected to impede molecular gas formation 
(e.g., Gerritsen \& de Blok 1999;
Mihos, Spaans, \& McGaugh 1999). Nonetheless, MW01
have shown that the edge-on LSB galaxy UGC~7321 contains a clumpy ISM
structure, including candidates for dark clouds, indicating that
it may have both the conditions and the catalysts for the formation
of H$_{2}$ and other molecules.

Despite this circumstantial evidence for molecular gas,
several groups have tried and failed to
detect CO emission from late-type LSB galaxies (Schombert et al. 1990;
Knezek 1993; de Blok \& van der Hulst 1998). 
However, these observations 
used only modest integration times, 
and are sufficient only to put
upper limits on their molecular gas contents of $M_{H2}/M_{HI}\le$0.07-0.43.
In fact, many of 
these values are consistent with the $M_{H2}/M_{HI}$ ratios typical of 
HSB Sd-Sm galaxies (Young \& Knezek 1989) and therefore
tell us little about how the molecular gas contents of LSB spirals may differ. 
Recently, O'Neil,
Hofner, \& Schinnerer (2000) reported deeper CO observations of 4 red
LSBs, including a detection of one object. However, the large \HI\ linewidth
of the detected galaxy ($W_{20}=430$~\kms) suggests it is  
one of the class of relatively rare LSB
giants (e.g., Sprayberry et al. 1995;
Matthews, van Driel, \& Monnier-Ragaigne 2001) and may
be somewhat different in nature from typical late-type LSBs.
Motivated by these factors,
we have undertaken  new deep CO(1-0) observations of 8 late-type,
low-mass, LSB spirals.

\section{Sample Selection}
For our survey we selected a sample of bulgeless, edge-on LSB galaxies
of types Scd-Sm. Targets of relatively large
angular size ($>$\am{2}{5}) were chosen so that the
telescope beam would be well matched to the expected 
CO-emitting regions in the inner disk.
We present here results for 8 galaxies with integration times $t\ga$5.5 hours.
Some general properties of our targets are summarized in Table~1. 

Although the {\it observed} surface brightnesses of our edge-on sample are
enhanced due to projection, their LSB natures are evident
from optical imaging and surface photometry (see Table~1).
All have a diffuse, transparent
appearance, lack a true dust lane, and have 
relatively high \dark\ ratios. In addition, all of these objects have
low star formation rates as evidenced by their 
weak or undetectable 1.4GHz radio
continuum fluxes in the NVSS survey (Condon et al. 1998).

We chose to emphasize edge-on LSB galaxies in the present study since 
projected gas surface densities are increased 
for edge-on viewing angles, hence the detectability of CO is
maximized. In addition,
edge-on systems provide us with a greatly enhanced ability to
identify dark
cloud candidates and obtain complementary measurements of 
the structure of the
dust and ionized gas (e.g., MW01).

\section{Data Acquisition and Reduction}
Our data were obtained with the NRAO\footnote{The
National Radio Astronomy Observatory (NRAO) is a facility of the
National Science Foundation operated under cooperative agreement by
Associated Universities, Inc.} 12-m Telescope at Kitt
Peak. Most observations were done 
in 1999 June, with some additional data acquired in 1997 January
and 1999 December. At 115.2GHz, 
the 12-m telescope has a beam of FWHP$\sim 55''$. 
Typical system temperatures during our runs were $T^{*}_{R}\sim$270--400K.

Our observations employed the 3mm dual SIS receivers and 
the two $256\times2$MHz filter bank spectrometers. 
Telescope pointing and
focus were checked frequently by observing planets or bright quasars.
Data were obtained in BSP (beam+position switching)
mode using the nutating secondary with a beam throw 
$\pm 2'$--$3'$. To eliminate spurious signals, we obtained roughly half of our
data with the source velocity offset $\pm$50~\kms\ from
the bandpass center. For all 8 targets we obtained data at the
optical center of the galaxy.
We also observed some offsets along 
the major axes of UGC~7321 and NGC~4244 (Table~2).

Our data were calibrated by performing a 
vane (chopper wheel) measurement after every other 6-minute
scan. Our absolute calibration was also checked by measuring
the CO line strength of M82 at least once during each observing
run.  We estimate an  absolute
flux accuracy of $\sim$20\%. 

We reduced our spectra using the CLASS and Unipops analysis packages.
All scans were individually inspected, and those
with highly structured baselines
were discarded. Scans were also examined
in blocks according to acquisition time  in order to
check for recurrent spurious signals. Channels repeatedly
exhibiting positive or
negative spikes $>3\sigma$ above the rms noise were blanked
and substituted with values interpolated from
the adjacent channels. Next a velocity window was chosen from 
\HI\ data, and a linear baseline was subtracted from
each scan. 
All remaining scans were then assigned weights according to their
mean rms noise and averaged to obtain the final spectra.

\section{Results}
The final spectra for each of our target galaxies are shown 
in Figure~1.  The data are
smoothed to a resolution of $\sim$20--40~\kms\ to enhance 
signal-to-noise. We detect the CO(1-0) line at $>4\sigma$ (after smoothing)
in 3 of our targets (UGC~2082, UGC~7321, NGC~4244),  
and marginally detect one object (UGC~6667) at $\sim3\sigma$.
These represent some of the
first  direct detections of molecular gas in late-type, LSB spiral galaxies.
Table~2 summarizes CO linewidths at 50\% peak maximum 
($W_{50,CO}$), line centroids ($V_{CO}$), and integrated CO line intensity
($I_{\rm CO}=\int T_R^* dV$), measured for the
detected galaxies using Gaussian fits. 

To within observational uncertainties we find that
the line centroids of our detected signals show good agreement
with \HI\ systemic velocities of these galaxies 
from the literature (see Table~1). 
In addition, the CO linewidths of our putative
detections are roughly comparable with the \HI\ linewidths.
This would be expected for CO spread over the
central few kpc of these LSB galaxies, as is generally found in late-type
spirals (e.g., Young \& Knezek 1989). 

One of our detected sources (NGC~4244) was previously 
detected in CO by Sage (1993a,b), offering corroborating evidence that
our detection is real. Like Sage (1993a,b)
we detected CO(1-0) emission both at the optical center of the galaxy as well
as at offset positions along the major axis. However, the strengths of
our measured lines are much weaker.
Since we used significantly longer integrations and the
12-m system performance is much improved since the late 1980's, our new
data strongly suggest that NGC~4244 is a weaker CO source than
previously reported.

For the 4 galaxies where no clear signal was detected, we
compute 3$\sigma$ upper limits using:
\begin{equation}
I_{CO}\le3T_{rms}\cdot W_{20,HI}/[f\times(1-W_{20,HI}/W)]^{0.5}
\end{equation}
where $T_{rms}$ is the rms noise in the final spectrum in mK,
$W_{20,HI}$ is the \HI\ linewidth, $f=W_{20,HI}/\delta_{c}$,
$\delta_{c}$ is
the channel spacing in \kms\, and $W$ is the entire velocity coverage
of the spectrum (Gao 1996). These
upper limits are summarized in Table~3.

\section{Discussion}
We estimate the total mass of molecular hydrogen within the 55$''$
telescope beam (Table 3) from:
\begin{equation}
M_{H2}=4.78[(\pi/(4\ln2)I_{CO}d^{2}_{b}]\epsilon^{-1}~~(M_{\odot})
\end{equation}
\noindent where $I_{CO}$ is the integrated CO line flux in
K$\cdot$\kms\ ($T^{*}_{R}$ scale),
$d^{2}_{b}$ is the telescope beam diameter in pc at
the distance of the galaxy, and $\epsilon$ is the main beam efficiency
of the telescope (0.84 at 115.3~GHz for the 12-m  as of 2000 January).
This formula assumes a Gaussian beam and a standard Galactic value of the
CO-to-H$_{2}$ conversion factor
$X=N(H_{2})/\int T(CO)dV=3.0\times10^{20}$cm$^{-2}$/[K \kms] 
(where $T(CO)=T^{*}_{R}/\epsilon$; e.g.,
Young \& Scoville 1991), which is applicable to the molecular clouds
at virial equilibrium. For galaxies where data were obtained at
off-center pointings, the total H$_{2}$ mass was
estimated by summing the individual spectra according to Eq.~3 of Sage (1993a).

The applicability of the 
standard $X$ factor to the diffuse, low metallicity environments
of LSB galaxies remains highly controversial (e.g., Maloney \& Black 1988;
Mihos et al.
1999), as low densities, low metallicities, and low dust contents
in LSBs would all seemingly
argue in favor a higher $X$ value, while clumpiness of
the ISM and/or the absence of a strong UV dissociating flux
may somewhat lower this number. This creates some uncertainty in our $M_{H2}$ 
estimates. Moreover, our derived CO fluxes 
should be regarded as conservative,
since data were obtained  at offset
pointings only for NGC~4244 and UGC~7321, and since some shadowing may
occur in edge-on galaxies if the CO is optically thick. 
However, this latter effect is expected to be small for LSB systems,
and based on the typical sizes of CO-emitting
regions in late-type galaxies, 
it is unlikely in the present survey that a significant fraction
of CO-emitting  gas lay outside our beam (see also below).

Previous null detections of CO in LSB galaxies have
been interpreted to suggest that either  CO does not trace 
H$_{2}$ in these galaxies or simply that
LSB galaxies are devoid of molecular  gas. 
It has even been postulated that stars may
form directly from atomic hydrogen in these galaxies (e.g., Schombert
et al. 1990). Our new observations show that at least some low-mass, 
late-type LSB galaxies do contain a modest molecular gas component.
If the Galactic $X$ factor is adopted, 
we find the implied $M_{H2}/M_{HI}$ ratios (0.010-0.055; Table~3) 
are several times smaller than those typically found for brighter Sd-Sm
spirals (0.19$\pm$0.10) by Young \& Knezek (1989). However, this does
not seem unexpected for dim, diffuse
galaxies that are inefficient at converting their raw, atomic gas into stars.

The strength of FIR (60 and 100$\mu$m) emission is well known
to correlate with CO strength in HSB galaxies (e.g., Sage 1993a), and
all 3
of our CO-detected LSBs were detected by {\it IRAS} (Table~1). We find these
galaxies follow the FIR-$M_{H2}$ correlation of Sage (1993a),
suggesting that their inferred molecular gas supplies are
adequate to sustain their observed star formation rates and that 
we have not grossly (i.e., by 
more than a factor of a few) underestimated their total H$_{2}$ contents.

In our present sample we find no other obvious
correlations between detection in CO and
global properties (e.g., color, \dark). This may simply reflect
our small sample size, and underscores the need for more molecular gas
observations of larger samples of LSB galaxies.

\section{Summary}
We have presented deep CO(1-0) observations of 8 edge-on, late-type,
LSB spiral galaxies. These are among the most sensitive CO
observations of LSB galaxies obtained to date. We have detected 3 of
our targets galaxies at a $>4\sigma$ level, and have marginally detected
one object at $\sim3\sigma$. These represent some of  the first direct
detections of molecular gas in low-mass, late-type LSB spiral galaxies.
Adopting a Galactic CO-to-H$_{2}$ conversion factor, we find 
values of $M_{H2}=1.4-3.5\times10^{7}$~\msun\ for the detected
galaxies, and upper limits 
$M_{H2}\la0.7-5.7\times10^{7}$~\msun\ for the undetected objects.
LSB spirals appear to be deficient in
molecular gas relative to their atomic gas contents compared with
brighter spirals of similar Hubble type, although the validity of the
Galactic CO-to-H$_{2}$ conversion factor for these systems is still
controversial.

\acknowledgements
We thank the observing staff at the 12-m Telescope, 
as well as W. van Driel and
D. Monnier-Ragaigne for their contributions in support of this project.
An anonymous referee also provided useful comments.
This research made use
of the NASA/IPAC Extragalactic Database (NED) which is operated by 
JPL under contract with NASA.



\begin{footnotesize}
\begin{deluxetable}{llcccccccccccl}
\tablewidth{46pc}
\tablenum{1}
\tablecaption{Properties of the Target Galaxies}

\tablehead{\colhead{Name} & \colhead{T} & \colhead{$V_{h}$} & 
\colhead{$D$} & \colhead{$i$} & \colhead{$D_{25}$} & 
\colhead{log$L_{B}$} & \colhead{$B-R$} & \colhead{$\mu(0)_{B,i}$} &
\colhead{$\bar\mu_{B,i}$} & \colhead{$W_{20,HI}$} & \colhead{log$M_{HI}$}
& \colhead{$L_{FIR}$} & \colhead{Ref.} \\
\colhead{} 
& \colhead{} & \colhead{(\kms)} &
\colhead{(Mpc)} & \colhead{($^{\circ}$)} & \colhead{($'$)} &
\colhead{($L_{\odot}$)} & \colhead{(mag)} & 
\multicolumn{2}{c}{(mag~$''^{-2}$)} & \colhead{(\kms)} &
\colhead{($M_{\odot}$)} &  \colhead{($10^{7}L_{\odot}$)} & \colhead{}
} 

\startdata


UGC711  & Sd & 1982 & 15.0 & 89 & 4.65 &
8.96 & 0.87& 24.4 & 25.8 & 230 & 9.04 & \nodata &  2,4,7 \nl

UGC2082 & Sc &  707 & 10.7  & 90 & 5.94 & 
9.10 & 1.34&  \nodata & 25.8 & 214 & 9.12 & 10.1 & 6,8\nl

UGC4148 & Sm &  737 & 12.8  & 90 & 2.52 & 
8.69 & \nodata &\nodata & \nodata & 148 & 8.68 & \nodata &  8 \nl

IC2233  & Sd &  563 & 10.0 & 90 & 5.17 & 
9.15 & 0.58& 23.1 & 24.8 & 206 & 9.08 & 7.0 & 4,8 \nl

UGC6667 & Scd&  979 & 15.5 & 90 & 3.70 & 
9.17 & 0.96&  23.8 & \nodata & 188 & 8.80 & \nodata & 9 \nl

NGC4244 & Scd & 244  & 3.6  & 84  &19.4 &
9.36 & 0.6: &  23.6:  & 24.4 & 180 & 9.13  & 13.4 & 1,5,6,8\nl

UGC7321 &  Sd  &  407 & 10.0 & 88  & 5.54 & 
9.00 & 0.99&  23.6 & 25.6   &  233 & 9.04 & 7.9  &  3 \nl

UGC9242 & Sd & 1440  & 20.6   & 90  & 5.66 & 
9.55  & 0.78&  23.7 & 25.7   & 205 & 9.29  & \nodata &  4,8 \nl

\enddata

\tablecomments{Explanation of columns: (1) name;  (2) Hubble type;
(3) heliocentric \HI\  recessional 
velocity; (4) distance; (5) inclination; (6) 
angular diameter; (7) logarithm of 
$B$-band luminosity; (8) $B-R$ color; (9) $B$-band central surface brightness
(corrected for inclination and internal
extinction); (10) mean $B$-band disk surface brightness within the outermost
observed isophote (corrected for inclination and internal extinction);
(11) \HI\ linewidth at 20\% peak maximum; (12) logarithm of the \HI\
mass; (13) FIR luminosity [see Helou et
al. (1988)];
(14) references for quoted optical and \HI\ parameters.
Type, $V_{h}$, $D_{25}$, and {\it IRAS} 60$\mu$m and 100$\mu$m fluxes
were taken from NED.  
The $B$-band central surface
brightness of NGC~4244 was estimated using Table~3 of  Ref.~1, 
and its global $B-R$ color was estimated from 
$B-R\approx1.5(B-V)+0.10$. If not quoted in the
original references, $B$-band
internal extinctions were estimated using Tully et al. (1998) and
surface brightnesses
were corrected to face-on  using Eq.~A4 
of Ref.~3 assuming $h_{r}/h_{z}\approx a/b$.}

\tablerefs{(1) Fry et al. (1999); (2) Gallagher et al., in prep.;
(3) Matthews et al.
(1999); (4) Matthews, Gallagher, \& van Driel
(2000); (5) Olling (1996); 
(6) Prugniel \& Heraudeau (1998); (7) Tifft \& Cocke (1988); 
(8) Tully (1988);  (9) Verheijen (1997).}

\end{deluxetable}
\end{footnotesize}

\begin{deluxetable}{lccccccc}
\tablewidth{45pc}
\tablenum{2}
\tablecaption{Summary of Observations}
\tablehead{\colhead{Name} & \colhead{$\alpha_{1950}$} &
\colhead{$\delta_{1950}$}  & \colhead{$t$} & \colhead{$T_{rms}$}
&\colhead{$T_{sys}$} &
 \colhead{$\delta_{c}$} & \colhead{Offset}\\
\colhead{} & \colhead{($^{h}~^{m}~^{s}$)} & \colhead{($^{\circ}$$~'~''$)} 
&  \colhead{(hours)} & \colhead{(mK)} & \colhead{(K)} 
& \colhead{(\kms)} & \colhead{(arcsec)}
}

\startdata
UGC711$^{a}$ & 010602 & +012229 & 5.5  & 0.94 & 525 & 41.6 & (0,0) \nl

UGC2082 &  023322 & +251222 & 7.8 & 0.67 & 399 & 41.6 & (0,0) \nl

UGC4148 & 075657 & +421953 & 6.0 & 1.05 & 340 & 20.8 & (0,0) \nl

IC2233  & 081027 & +455343 & 7.8 & 0.72 & 317 & 20.8 & (0,0) \nl

UGC6667 &  113945 & +515232 & 8.9 & 0.68 & 354 & 41.6 & (0,0) \nl

NGC4244a &121500 & +380509 &  7.2 & 1.10 & 325 & 20.8 & (0,0) \nl

NGC4244b& $''$ & $''$ & 1.6 & 2.37 & 358 & 20.8 & ($-$21.9,$-$20.)\nl

NGC4244c& $''$ & $''$ & 5.1 & 1.15 & 319 & 20.8 & (+45.,+40.) \nl

NGC4244d& $''$ & $''$ & 7.0 & 1.22 & 332 & 20.8 & ($-$46.,$-$30.) \nl

NGC4244e& $''$ & $''$ &  2.0 & 2.03 & 316 & 20.8 & (+46.,+30.) \nl

UGC7321a & 121503 & +224903&  15.0 & 0.47 & 340 & 20.8 & (0,0) \nl

UGC7321b & $''$ & $''$ &  13.4 & 0.60 & 308 & 20.8 & ($-$27.,$-$3.8) \nl

UGC9242 &  142319 & +394550 & 5.6 & 0.98 & 330 & 41.6 & (0,0) \nl

\enddata
\tablenotetext{a}{Receiver 2 exhibited anomalously high
$T_{sys}$ during observations of UGC~711.}
\tablecomments{Explanation of columns: (1) galaxy name; (2) \& (3)
B1950.0 coordinates (4) total
usable integration time; (5) spectrum rms; (6) mean system
temperature of all averaged scans; (7) velocity width of
channels in final spectrum, after Hanning smoothing; (8)
coordinates of observations, in units of arcseconds north and east from
the target's optical center. }

\end{deluxetable}

\begin{deluxetable}{llllccc}
\tablenum{3}
\tablecaption{Measured  CO Parameters}

\tablehead{\colhead{Name}  & \colhead{$V_{CO}$} &
\colhead{$W_{50,CO}$} & 
\colhead{$I_{CO}$} & \colhead{S/N} &
\colhead{$M_{H2}$} & \colhead{$M_{H2}/M_{HI}$} \\
\colhead{} &  \colhead{(\kms)} 
& \colhead{(\kms)} &
\colhead{(K$\cdot$\kms)}  & \colhead{} & 
\colhead{($10^{7} M_{\odot}$)} & \colhead{} }
 
\startdata

UGC711 & \nodata & \nodata & $\le$0.30 & \nodata & $\le$3.14 & $\le$0.028 \nl

UGC2082 & 725.9 (15.0) & 176.9 (26.4) & 0.57 (0.09) & 4.7 & 2.98 &
0.022 \nl

UGC4148 & \nodata &  \nodata &  $\le$0.17 & \nodata & $\le$1.24 & 
$\le$0.026 \nl

IC2233  & \nodata &  \nodata & $\le$0.15 & \nodata & $\le$0.71 & 
$\le$0.0058 \nl

UGC6667 &  971.4: (15.0) & 161.7: (44.3) & 0.32: (0.08) & 2.9 & 3.52 &
0.055 \nl

NGC4244a &   217.1 (8.7) & 194.4 (24.6) &
0.91 (0.08) & 6.3 & 0.54 & \nodata \nl

NGC4244b&   213.4: (31.7) & 223.9: (57.2) & 0.75: (0.20) & 2.9 & 0.44
& \nodata \nl

NGC4244c&   193.5 (8.1) & 126.0 (21.5) & 0.85 (0.10) & 5.8 & 0.50 &
\nodata \nl

NGC4244d&   249.0 (17.0) & 89.6 (41.9) &
0.70 (0.22) & 4.7 & 0.42 & \nodata \nl

NGC4244e& 211.2 (5.7) & 51.5 (11.6) & 0.43 (0.09) & 3.8 & 0.26 & \nodata \nl

NGC4244(sum) & 227.4 (5.9) & 136.6 (12.7) & 2.32 & \nodata & 1.36 & 0.010\nl

UGC7321a & 385.9 (10.0) & 168.2 (25.7) & 0.50 (0.06) & 8.2 & 2.30 &
\nodata \nl

UGC7321b & 469.5 (8.6) & 75.0 (20.1) & 0.20 (0.04) & 4.5 & 0.92 &
\nodata \nl 

UGC7321(sum) & 420.5 (10.9) & 190.3 (21.2)  & 0.71  & \nodata  & 3.22
& 0.029 \nl

UGC9242 & \nodata & \nodata & $\le$0.30 & \nodata & $\le$5.75 & $\le$0.029 \nl

\enddata

\tablecomments{CO line parameters are measured
from Gaussian fits; upper limits are derived as described in Sect.~4. 
Uncertainties in measured quantities
are given in parentheses following the entry.
A colon (:) following a table entry
indicates a marginal ($\sim3\sigma$) detection. 
Explanation of columns: (1) name; (2) velocity centroid
of CO(1-0) line; (3) 
full width at 50\% peak maximum of the
CO(1-0) line; (4)  integrated
CO(1-0) line flux ($T^{*}_{R}$ scale) 
derived from a Gaussian fit; upper limits are given
for cases where no signal was detected at $\ga3\sigma$; fluxes may be
converted to Jy by multiplying by the conversion factor $f_{1}\approx$35~Jy/K;
(5) signal-to-noise of the detected CO line; (6)
estimate of the mass of molecular hydrogen within the 55$''$ FWHP 
beam (see Sect.~5);
(7) ratio of the estimated amount of
molecular hydrogen to the total \HI\ mass of the galaxy. Entries
denoted ``sum'' are derived by summing observations at multiple
pointings (see Sect.~5.)}

\end{deluxetable}


\begin{figure}[t]
\plotfiddle{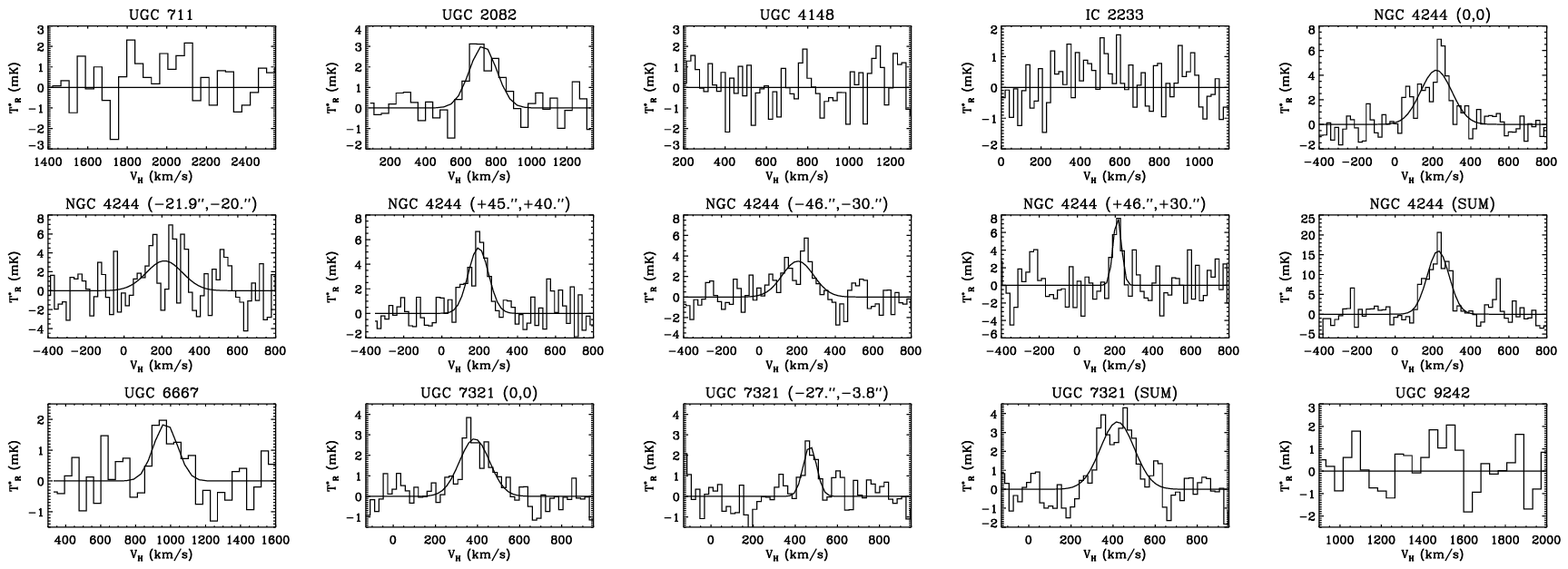}{2.5in}{0}{100}{100}{-320}{-400}
\caption{CO(1-0) spectra 
of 8 edge-on, late-type, low surface
brightness galaxies obtained with the NRAO 12-m telescope. 
Axes are radial velocity (in \kms) versus CO
antenna brightness temperature in mK ($T^{*}_{R}$ scale). 
Overplotted lines show Gaussian fits if a signal was detected. The
spectra marked `SUM' show the combined results of multiple pointing
observations.}
\end{figure}


\begin{references}
%

Condon, J. J. et al. 1998, AJ, 115, 1693

de Blok, W. J. G. \& van der Hulst, J. M. 1998, A\&A, 336, 49


Fry, A. M., Morrison, H. L., Harding, P., \& Boroson, T. A. 1999, AJ,
118, 1209

Gao, Y. 1996, Ph.D. Thesis, State Univ. of New York at Stony Brook

Gerritsen, J. P. E. \& de Blok, W. J. G. 1999, A\&A, 342, 655

Helou, G., Khan, I. R., Malek, L., \& Boehmer, L. 1988, ApJS, 68, 151



Knezek, P. M. 1993, Ph.D. Thesis, Univ. of Massachusetts


Maloney, P. \& Black, J. H. 1988, ApJ, 325, 389

Matthews, L. D. \& Gallagher, J. S. 1997, AJ, 114, 1899

Matthews, L. D., Gallagher, J. S., \& van Driel, W. 1999,  AJ, 118,
2751

Matthews, L. D., Gallagher, J. S., \& van Driel, W. 2000, in Galaxy
Dynamics: from the Early Universe to the Present, ed. R
F. Combes, G. A. Mamon, \& V. Charmandaris, (San Francisco: ASP), 195

Matthews, L. D., van Driel, W., \& Monnier-Ragaigne, D. 2001, A\&A,
365, 1

Matthews, L. D. \& Wood, K. 2001, ApJ, 548, in press (MW01)

McGaugh, S. S. 1994, ApJ, 426, 135

McGaugh, S. S. \& Bothun, G. D. 1994, AJ, 107, 530

Mihos, J. C., Spaans, M., \& McGaugh, S. S. 1999, ApJ, 515, 89

Olling, R. P. 1996, AJ, 112, 457

O'Neil, K., Hofner, P., \& Schinnerer, E. 2000, ApJ, 545, L99

Prugniel, P. \& Heraudeau, P. 1998, A\&AS, 128, 299


Sage, L. J. 1993a, A\&A, 272, 123

Sage, L. J. 1993b, A\&AS, 100, 537

Sanders, D. B. et al. 1986, ApJ, 305, 45

Schombert, J. M., Bothun, G. D., Impey, C. D., \& Mundy, L. G. 1990,
AJ, 100, 1523

Schombert, J. M., Bothun, G. D., Schneider, S. E., \& McGaugh,
S. S. 1992, AJ, 103, 1107

Sprayberry, D., Impey, C. D., Bothun, G. D., \& Irwin, M. J. 1995, AJ,
109, 558


Tifft, W. G. \& Cocke, W. J. 1988, ApJS, 67, 1

Tully, R. B. 1988, Nearby Galaxies Catalog (Cambridge: Cambridge
University Press)

Tully, R. B. Pierce, M. J., Huang, J.-S., Saunders, W., Verheijen,
M. A. W., \& Witchalls, P. L. 1998, AJ, 115, 2264

van der Hulst, J. M., Skillman, E. D., Smith, T. R., Bothun, G. D.,
McGaugh, S. S., \& Bothun, G. D. 1993, AJ, 106, 548


Verheijen, M. A. W. 1997, Ph.D. Thesis, University of Groningen

Young, J. S. \& Knezek, P. M. 1989, ApJ, 347, L55

Young, J. S., \& Scoville, N. Z. 1991, ARA\&A, 29, 581

\end{references}
\end{document}